\documentclass[12pt]{article}

\usepackage{epsf}

\def\theequation{\thesection.\arabic{equation}}
\newcommand{\ssection}[1]{
\section{#1}
\setcounter{equation}{0}
}

\def\np#1#2#3{{\it Nucl. Phys.} {\bf B#1} (#2) #3}

\def\physrev#1#2#3{{\it Phys. Rev.} {\bf D#1} (#2) #3}

\def\physrep#1#2#3{{\it Phys. Rep.} {\bf #1} (#2) #3}

\def\cmp#1#2#3{{\it Comm. Math. Phys.} {\bf #1} (#2) #3}
\def\mpl#1#2#3{{\it Mod. Phys. Lett. }{\bf #1} (#2) #3}
\def\ijmp#1#2#3{{\it Int. J. Mod. Phys.} {\bf #1} (#2) #3}

\def\jmp#1#2#3{{\it J. Math. Phys.} {\bf #1} (#2) #3}
\def\prsl#1#2#3{{\it Proc. Roy. Soc. London} {\bf #1} (#2) #3}
\def\jhep#1#2#3{{\it JHEP} {\bf #1} (#2) #3}
\def\hepth#1{{\tt hep-th/}#1}

\def\Ref#1{(\ref{#1})}

\def\subsub#1{
\vskip 0.2cm

{\it #1}
}

\newcommand{\be}{\begin{equation}}
\newcommand{\ee}{\end{equation}}
\newcommand{\beq}{\begin{eqnarray}}
\newcommand{\eeq}{\end{eqnarray}}
\newcommand{\bea}[2]{\be\label{#2}\begin{array}{#1}}
\newcommand{\eea}{\end{array}\ee}

\def\lfig#1#2#3#4{
 \begin{figure}
 \refstepcounter{figure}
 \label{#4}
 \addtocounter{figure}{-1}
 \epsfxsize=#3
 \centerline{\epsfbox{#2}}
 {\bf \caption{{\rm #1}}}
 \end{figure}
}

\def\Rb{{\rm \bf R}}

\def\Zb{{\rm \bf Z}}
\def\Tr{\,{\rm Tr}\, }

\def\diag{{\rm diag}}

\def\({\left(}
\def\){\right)}
\def\[{\left[}
\def\]{\right]}
\def\p{\partial}

\def\11{1\!\! 1}

\def\hf{{1\over 2}}

\def\e{\epsilon}

\def\s{\sigma}

\def\o{\omega }
\def\dd{\nabla}

   \def\CF {{\cal F}}

   \def\CN {{\cal N}}

   \def\CR {{\cal R}}

\def\op{\omega_{_{+}}}

\def\tp{t_1}
\def\tm{t_{-1}}

\def\Xp{X_{+}}
\def\Xm{X_{-}}
\def\Xpm{X_{\pm }}

\def\bff{\varphi}
\def\ff{\varphi_0}
\def\dl{\Phi}
\def\ddl{\phi}
\def\mt{m_{\eta}}

\def\os{\o_{\rm sing}}
\def\op{\tilde\o}
\def\pp{\tilde p}
\def\xip{\tilde \xi}
\def\xtp{\tilde x_t}
\def\xs{x_{\rm sing}}
\def\tf{\tau}
\def\xf{q}
\def\xt{x_t}

\def\dd{\nabla}
\def\kR{{k\over R}}
\def\oR{{1\over R}}
\def\alp{\alpha'}
\def\pal{\alpha'^{-1}}

\def\nspal#1{{#1\over \sqrt{\alp}}}

\def\xt{\xi}

\begin{document}

\renewcommand{\thefootnote}{$\fnsymbol{footnote}$}

\

\vskip 4cm

\begin{center}
\Large \bf
Backgrounds of 2D string theory from matrix model
\end{center}

\

\centerline{ Sergei Alexandrov}

\

\begin{center}
\it Service de Physique Th\'eorique,
CNRS - URA 2306, C.E.A. - Saclay, \\
F-91191 Gif-Sur-Yvette CEDEX, France \\
Laboratoire de Physique Th\'eorique de l'\'Ecole Normale
Sup\'erieure\footnote[2]{Unit\'e de Recherche du Centre
National de la Recherche Scientifique et de  l'Ecole Normale
Sup\'erieure et \`a l'Universit\'e de Paris-Sud.}\\
24 rue Lhomond, 75231 Paris Cedex 05, France \\
V.A.~Fock Department of Theoretical Physics,
St.~Petersburg University, Russia
\end{center}

\begin{abstract}
In the Matrix Quantum Mechanical formulation of 2D string theory it is
possible to introduce arbitrary tachyonic perturbations.  In the case
when the tachyonic momenta form a lattice, the theory is known to be
integrable and, therefore, it can be used to describe the
corresponding string theory. We study the backgrounds of string
theory obtained from these matrix model solutions.
They are found to be flat
but the perturbations can change the  global structure of
the target space. They can lead either to a compactification, or to
the presence of boundaries depending on the choice of boundary conditions.
Thus, we argue that the tachyon perturbations have a dual
description in terms of the unperturbed theory in spacetime with
a non-trivial global structure.
\end{abstract}

\newpage

\renewcommand{\thefootnote}{\arabic{footnote}}

    \ssection{Introduction}

The two-dimensional string theory has proven to be a model
possessing a very rich structure
\cite{metr:KLEBANOV,metr:Moore,metr:JEVICKI}.
It allows to address various questions inaccessible in higher
dimensional string theories as well as it appears as an equivalent
description of topological strings and supersymmetric gauge
theories \cite{metr:MUKHIVAFA,metr:GOSHALVAFA,metr:DVd}.
Perhaps, its main feature is the integrability. Many problems
formulated in its framework, such as calculation of $S$-matrix,
were found to be exactly solvable.

However, this integrability is hidden in the usual CFT formulation
and becomes explicit only when one uses a matrix model description
of string theory.
In this case the model, one has to consider, is the Matrix Quantum
Mechanics (MQM) \cite{metr:KAZMIG}.
It was successfully used to describe 2D string theory
in the simplest linear dilaton background as well as to incorporate
perturbations. In two dimensions there are basically two types of
possible perturbations: tachyon and winding modes. The former
always present in the theory as momentum modes of a massless field
which, following tradition, is called ``tachyon''.
The latter appear when one considers Euclidean theory compactified
on a time circle.

There are two ways to change the background of string theory:
either to consider strings propagating in a non-trivial target space
or to introduce the perturbations mentioned above. In the first case
one arrives at a complicated sigma-model. Not many examples
are known when such a model turns out to be solvable. Besides,
it is extremely difficult to construct a matrix model realization
of a general sigma-model since not much known about matrix
operators explicitly perturbing the metric of the target space.
Thus, we lose the possibility to use the powerful
matrix model machinery to tackle our problems.

On the other hand, following the second way, we find that
the integrability of the theory in the trivial background
is preserved by the perturbations. Namely, when we restrict our attention
only to windings or to tachyons with discrete momenta, they
are described by Toda integrable hierarchy
\cite{metr:DMP,metr:KKK,metr:AKK}.
Therefore, the backgrounds of string theory obtained by these
perturbations represent a good laboratory for investigation
of different problems.

One of the such interesting problems is the black hole thermodynamics.
As it is well known, 2D string theory possesses an exact background
which incorporates the black hole singularity \cite{metr:MSWbh,metr:WITbh}.
From the target space point of view, it is described just by one
of the dilaton models, the so-called CHGS model \cite{metr:CGHS}. Its
thermodynamical properties have been extensively studied
(see, for example, the review on dilaton models \cite{metr:DilRev}).
However, to achieve the microscopical statistical description,
one should deal with the full string theory
so that one has to obtain the complete solution of the corresponding
sigma-model.

Fortunately, there exists an alternative description of string theory in
this background. It has been conjectured
that the black hole background can be created also by introduction
of the winding modes \cite{metr:FZZ}. Moreover, a matrix model incorporating
these perturbations has been proposed \cite{metr:KKK}.
Thus, we get a possibility
to describe dynamics of string theory in non-trivial backgrounds
using the usual perturbations of MQM.

In this paper we consider tachyon perturbations of 2D string theory
which are T-dual to perturbations by winding modes.
We saw that the latter are
supposed to create curvature of the target space.
Do tachyon perturbations can also change the structure of
the target space? The paper is devoted to addressing this question.

First of all, we explain why we study tachyon and not
winding perturbations, whereas these are the windings that have
the most interesting physical applications like the physics of black holes.
The main reason is that it is much easier to study the tachyon modes
from the MQM point of view.
In this case we have a powerful description in terms of free fermions
on the one hand \cite{metr:BIPZ} and
the Das--Jevicki collective field theory on the other \cite{metr:DasJev}.
The latter has a direct target space interpretation what allows to identify
matrix model quantities with characteristics of the string background.
However, this description is absent when the winding modes are
included what is possibly related to the absence of a local target space
interpretation of windings.

Tachyon modes do have such local interpretation and, therefore,
the background, which they create, is characterized simply by
a (time-dependent) condensate of these modes.
However, we are looking for another description which would be dual
to the one with the tachyon condensate. We expect it to be
similar to the dual description of the winding perturbations,
that is it should replace the time-dependent tachyon condensate by
a change of the structure of the target space.

As one of justifications of why such dual description should exist,
we mention that it has been shown that the tachyon perturbations lead to
a finite temperature as if we had a black hole \cite{metr:AKTBH}.
In fact, to give rise a temperature, it is not necessary
for the target space to contain a black hole singularity.
The thermal description can arise in very different situations
including flat spacetimes, but in any case it should be different
from the usual Minkowski space \cite{metr:BirDav}.

In this paper we show that comparison of the Das--Jevicki collective
theory with the low-energy effective action for the string background
leads to the picture we sketched above. Instead of the tachyon
condensate we get the target space with a non-trivial global structure.
Non-triviality can appear either as boundaries or as a compactification.
Locally spacetime remains the same as it was before the perturbation.

The paper is organized as follows. In the next section we briefly
review 2D string theory in the CFT formulation. In section 3
the Das--Jevicki collective field theory for the singlet sector of MQM
is analyzed. We derive an effective action for the quantum correction to
the eigenvalue density of MQM. In section 4 this correction
and its action are identified
with the tachyon field of string theory and the low-energy
effective action for the string background, correspondingly.
The identification imposes
a condition which leaves us with the linear dilaton background.
However, there is still a freedom to have a non-trivial global structure
which is investigated in section 5. The case of the simplest
Sine--Liouville perturbation is considered in detail.

\ssection{Tachyon and winding modes in 2D Euclidean string theory}

2D string theory is defined by the Polyakov action
\be\label{metr:PSTR}
S_{\rm P}={1\over 4\pi}\int d^2\sigma\sqrt{h}
[ h^{ab}\p_a X\p_b X +\mu],
\ee
where the bosonic field $X(\sigma)$ describes the embedding of the
string into the Euclidean time dimension and $ h^{ab}$ is a world
sheet metric.  In the conformal gauge $h^{ab}=e^{\phi(\s)}\hat h^{ab}$,
where $\hat h^{ab}$ is a background metric, the conformal mode
$\phi$ becomes dynamical due to the conformal anomaly and the
world-sheet CFT action takes the familiar Liouville form
\be\label{metr:confstr}
S_0={1\over4\pi}\int d^2 \sigma\, [(\partial X)^2
+(\partial\phi)^2 -2\hat \CR\phi + \mu e^{-2\phi}+{\rm ghosts}].
\ee
This action is known to describe the unperturbed linear dilaton string
background corresponding to the flat 2D target space parameterized
by the coordinates $(X,\phi)$ \cite{metr:POLCHINSKI}.

In two dimensions string theory possesses only one field-theoretic
degree of freedom, which is called ``tachyon'' since it corresponds
to the tachyon mode of the bosonic string in 26 dimensions.
However, in the present case it is massless.
The vertex operators for the tachyons of momentum $p$ are written as
\be\label{metr:vert}
V_p = {\Gamma(|p|)\over \Gamma(-|p|)}\int d^2
 \sigma\, e^{-i p X }e^{( |p|-2)\phi}.
\ee
If we consider the Euclidean theory compactified to a time circle
of radius $R$, the momentum is allowed to take only discrete values
$p_n=n/R$, $n\in \Zb$. In this case there are also winding modes
of the string around the compactified dimension which also have
a representation in the CFT \Ref{metr:confstr}
as vortex operators $\tilde V_q$.

Both tachyons and windings can be used to perturb the simplest
theory \Ref{metr:confstr}
\be\label{metr:PERTS}
 S=S_0+\sum_{n\ne 0}( t_n V_n +\tilde t_n\tilde V_n).
\ee
How do these perturbations look from the target space point of
view? The presence of non-vanishing couplings $t_n$ (together with
the cosmological constant $\mu$) means that we consider a string
propagation in the background with a non-trivial vacuum expectation
value of the tachyon. In particular, the introduction of momentum
modes makes it to be time-dependent. On the other hand, the winding
modes, which are defined
in terms of the T-dual world sheet field $\tilde X=X_R-X_L$, where
$X_R$ and $X_L$ are the right and left-moving components of the
initial field $X(\sigma)$,
have no obvious interpretation in terms of massless background fields.
Since windings are related to global properties of strings,
they should correspond to some non-local quantities in the low-energy
limit. Their condensation can result in change of the background
we started with.

A particular proposal for the structure of the resulting background
has been made for the simplest case of $\tilde t_{\pm 1}\ne 0$
\cite{metr:FZZ}.
The CFT with the lowest winding perturbation has been conjectured to
be dual to the WZW ${\rm SL(2,\Rb)/U(1)}$ sigma-model, which describes
the 2D string theory on the black hole background
\cite{metr:MSWbh,metr:WITbh,metr:DVV}.
In \cite{metr:AKK} it was suggested that any perturbation of type
 \Ref{metr:PERTS} should have a dual description as string theory in
 some non-trivial background, which is determined by the two sets of
 couplings $t_n$ and $\tilde t_n$.  Thus, not only for winding, but
 also for tachyon perturbations we expect to find such a description.

Note, that the two sets of perturbations are related by T-duality
on the world sheet \cite{metr:KLEBANOV}. However, this does
not mean that the corresponding string backgrounds should be the same.
This is because the momentum and winding modes have different
target space interpretations and it is not clear how to
carry over this T-duality from the world sheet to the target space.

\ssection{Das--Jevicki collective field theory}

The 2D string theory can be described using Matrix Quantum Mechanics
in the double scaling limit (for review see\cite{metr:KLEBANOV}).
The matrix model Lagrangian corresponding to the unperturbed CFT
\Ref{metr:confstr} is
\be\label{metr:mmL}
L=\hf \Tr\left({\dot M}^2-u(M)\right),
\ee
where $u(M)=-\hf M^2$ is the inverse oscillator potential.
It is known also how to introduce tachyons and windings into this model
\cite{metr:AKK,metr:GRKL,metr:BULKA,metr:KKK}.
In the following we restrict ourselves to the tachyon perturbations.
In this case it is enough to consider only the singlet sector of MQM,
where the wave function is a completely antisymmetric function
of the matrix eigenvalues $x_i$. Due to this the system has an equivalent
description in terms of free fermions in the potential $u(M)$.

The fermionic description is very powerful and allows to calculate
many interesting quantities in the theory.
However, to make contact with the target space
description of string theory, another representation has proven
to be quite useful. It is obtained by a bosonization procedure and leads
to a collective field theory developed by Das and
Jevicki \cite{metr:DasJev,metr:JEVICKI}.
The role of the collective field is played by the density of
matrix eigenvalues
\be\label{metr:cmf}
\bff(x,t)=\Tr\delta\left(x-M(t)\right).
\ee
If to introduce the conjugated field $\Pi(x)$ such that
\be\label{metr:imp}
\{ \bff(x), \Pi(y)\}=\delta(x,y),
\ee
the effective action for the collective theory is written as
\beq
S_{\rm coll}&=\int dt \left[\int dx\,
\left(  \Pi \p_t \bff\right) -H_{\rm coll}\right],
\label{metr:Scolm}\\
{\rm where}\quad H_{\rm coll}&=\int dx \left(
{1\over 2} \bff(\p_x\Pi)^2 +{\pi^2 \over 6}\bff^3
+(u(x)+\mu)\bff \right) \nonumber
\eeq
and we added the chemical potential $\mu$ to the action.
The classical equations of motion read
\beq
-\p_x \Pi&=&{1\over \bff}\int dx \p_t \bff, \label{metr:cleq} \\
-\p_t \Pi &=&
\hf (\p_x \Pi)^2 +{\pi^2\over 2} \bff^2 + (u(x)+\mu).
\eeq
Due to the first equation one can exclude the momenta and pass to
the Lagrangian form of the action
\be\label{metr:Scol}
S_{\rm coll}=\int dt \int dx\,
\left({1\over 2\bff}\left(\int dx \p_t\bff\right)^2-{\pi^2 \over 6}\bff^3
-(u(x)+\mu)\bff\right)
\ee
with the following equation for the classical field
\be\label{metr:Leq}
\p_t\left( {\int dx \p_t \bff \over \bff} \right)=
\hf \p_x \left( {\int dx \p_t \bff \over \bff} \right)^2
+{\pi^2\over 2} \p_x \bff^2 +\p_x u(x).
\ee

Let us consider a function ${1\over \pi}\ff(x,t)$,
which is a solution of the classical equation of motion \Ref{metr:Leq}.
We expand the collective field around the background given by
this solution
\be\label{metr:expf}
\bff(x,t)={1\over \pi}\ff(x,t)+{1\over \sqrt{\pi}}\p_x\eta(x,t).
\ee
The action takes the form
\be\label{metr:expS}
S_{\rm coll}=S_{(0)}+S_{(1)}+S_{(2)}+\cdots,
\ee
where $S_{(0)}$ is a constant, $S_{(1)}$ vanishes due to the
equation \Ref{metr:Leq}, and $S_{(2)}$ is given by
\be\label{metr:Sd}
S_{(2)}=\hf \int dt \int {dx\over \ff}\, \left[
(\p_t \eta)^2 -2  {\int dx \p_t \ff \over \ff}\p_t\eta \p_x \eta
-\left( \ff^2 -\left( {\int dx \p_t \ff \over \ff}\right)^2
\right) (\p_x\eta)^2  \right].
\ee

We want to use the quadratic part of the action to extract
information about the background of string theory corresponding
to the solution $\ff(x,t)$. For that we will identify
the quantum correction $\eta$ with the tachyon field and
interpret \Ref{metr:Sd} as the kinetic term for the tachyon in the
low-energy string effective action. Therefore, we will need
the properties of the matrix standing in front of derivatives of $\eta$.
The crucial property is that for any $\ff$ its determinant is equal to $-1$.
As a result, the action can be represented in the following form
\be\label{metr:Sg}
S_{(2)}=-\hf \int dt \int dx\, \sqrt{-g} g^{\mu\nu}\p_\mu \eta \p_\nu\eta.
\ee
In a more general case we would have to introduce a dilaton dependent
factor coupled with the kinetic term.

The action \Ref{metr:Sg} is conformal invariant. Therefore, one can always choose
coordinates where the metric takes the usual Minkowski form
$\eta_{\mu\nu}=\diag(-1,1)$.
Unfortunately, we do not know the explicit form of this transformation
for arbitrary solution $\ff$. Note, however, that one can bring
the metric to another standard form which is of the Schwarzschild type.
Indeed, it is easy to check that if one changes the $x$ coordinate to
\be\label{metr:tap}
y(x,t)=\int \ff(x,t) dx,
\ee
the metric becomes
\be\label{metr:met}
g_{\mu\nu}=
\pmatrix{-\ff^2 & 0 \cr
           0 & \ff^{-2}}.
\ee

\ssection{Low-energy effective action and background of 2D string theory}

Now we turn to the string theory side of the problem.
In a general background the string theory is defined by the following
$\sigma$-model action
\be\label{metr:smod}
S_\sigma={1\over4\pi\alp}\int d^2 \sigma\,\sqrt{h}
\left[G_{\mu\nu}(X)h^{ab}\p_a X^{\mu} \p_b X^{\nu}
+\alp\hat\CR \dl(X) + T(X)\right],
\ee
where $X(\sigma)$ is a two-dimensional field on the world sheet.
As it is well known the condition that the conformal invariance of
this theory is unbroken reduces to some equations on
the background fields: target space metric $G_{\mu\nu}$,
dilaton $\dl$ and tachyon $T$.
These equations can be obtained from the effective action which in
the leading order in $\alpha'$ is given by \cite{metr:Polbook}
\be\label{metr:Seff}
S_{\rm eff}=\hf \int d^2 X \, \sqrt{-G}e^{-2\dl } \left[ {16\over \alp}+
R+4(\dd\dl)^2-(\dd T)^2+{4\over\alp}T^2\right].
\ee
Here the covariant derivative
$\dd_{\mu}$ is defined with respect to the metric $G_{\mu\nu}$ and $R$
is its curvature. The first term comes from the central charge and in
$D$ dimensions looks as ${2(26-D)\over 3\alp}$ disappearing in the
critical case.
The equations of motion following from the
action \Ref{metr:Seff} are equivalent to
\beq
\beta_{\mu\nu}^G &=&R_{\mu\nu}+2\dd_{\mu}\dd_{\nu} \dl
-\dd_{\mu}T \dd_{\nu}T=0,
\nonumber \\
{4\over\alp}\beta^{\dl} &=&-{16\over\alp}
-R+4(\dd\dl)^2-4\dd^2\dl+(\dd T)^2-{4\over\alp}T^2=0,
\label{metr:eqss} \\
\beta^T &=&-2\dd^2 T +4 \dd\dl\cdot \dd T-{8\over\alp}T=0. \nonumber
\eeq

Let us concentrate on the tachyonic part of the action \Ref{metr:Seff}.
We want to relate it to the action \Ref{metr:Sg} for the collective theory obtained
form Matrix Quantum Mechanics. In particular, the tachyon field
$T$ and quantum fluctuation of density $\eta$
should be correctly identified with each other.
We see that the main difference between \Ref{metr:Seff}
and \Ref{metr:Sg} is the presence of the dilaton dependent factor $e^{-2\dl}$.
To remove this factor, we redefine the tachyon as $T=e^{\dl}\eta$.
Then the part of the action quadratic in the tachyon field reads
\be\label{metr:Stach}
S_{\rm tach}=-\hf \int d^2 X\,
\sqrt{-G} \left[(\dd\eta)^2+\mt^2 \eta^2\right],
\ee
where
\be\label{metr:mas}
\mt^2=(\dd\dl)^2-\dd^2\dl-{4\pal}.
\ee
We denoted the new tachyon by the same letter as the matrix model field since
they should be identical. However, comparing two actions \Ref{metr:Stach}
and \Ref{metr:Sg}, we see that for this identification to be true,
one must require the vanishing of the tachyonic mass
\be\label{metr:vmas}
\mt^2=0.
\ee
This equation plays the role of an additional condition for
the equations of motion \Ref{metr:eqss}
and allows to select the unique solution.
It is easy to check that the linear dilaton background
\be\label{metr:solld}
G_{\mu\nu}=\eta_{\mu\nu}, \qquad \dl=-\nspal{2} X^1,
\qquad \eta=0
\ee
satisfies all these equations.
Besides, we argue that it is the unique solution of
the equations \Ref{metr:eqss} and \Ref{metr:vmas}.
It means that for any solution of MQM restricted to
the singlet sector we always get the simplest linear dilaton background.
Since the theory \Ref{metr:Seff} is a model of two-dimensional dilaton
gravity with non-minimal coupled scalar field, it is not exactly
solvable \cite{metr:DilRev}. Therefore, we are not able to prove our statement
rigorously. However, in Appendix A we show that there are no solutions
which can be represented as an expansion around the linear dilaton
background \Ref{metr:solld}.\footnote{Of course, this result is valid only for
the equations of motion obtained in the leading order in $\alp$.
For example, the background described by the CFT \Ref{metr:confstr} or even
\Ref{metr:PERTS} with $\tilde t_n=0$ corresponds
to the linear dilaton background
with a non-vanishing tachyon field. Such background
does not satisfy the equations \Ref{metr:eqss}.
However, it was suggested that it satisfies the exact equations
which include all corrections in $\alp$ \cite{metr:Moore}. We are interested
in the existence of a dual description where the non-vanishing tachyon
vacuum transforms into expectation values of other fields.
Our results show that the curvature of the target space and the dilaton
can not be changed in the leading order in $\alp$.}

Thus, any tachyon perturbation can not modify the local structure
of spacetime. However, the found solution does not forbid modifications
of the global structure. There are, in principle, two
possibilities to change it. Either something can happen with
the topology, for example, a spontaneous compactification,
or there can appear boundaries.
We argue that just these possibilities are indeed realized in our
case.

To understand why this can happen, note that in the initial
coordinates $(t,x)$ of MQM the metric is non-trivial, although flat.
We can represent it as
\be\label{metr:mets}
G_{\mu\nu}=e^{2\rho}g_{\mu\nu} ,
\ee
where $g_{\mu\nu}$ is the matrix coupled with the derivatives
of $\eta$ in \Ref{metr:Sd}, whose determinant equals $-1$.
It takes the usual form $\eta_{\mu\nu}$ only after
a coordinate transformation.
Under this transformation, the image of the $(t,x)$-plane is, in general,
not the whole plane, but some its subspace. Depending on boundary
conditions, it will give rise either to a compactification, or to
the appearance of boundaries.

\ssection{Target space of string theory with
integrable tachyon perturbations}

In this section we analyze the structure of the target space of
2D string theory in the presence of tachyon perturbations with
discrete momenta.
Such perturbations correspond to the admissible perturbations
of the Euclidean theory described in section 2. Nevertheless, our analysis
is carried out in spacetime of the Minkowskian signature.

The case, we consider here, was shown to be exactly integrable
and described by Toda Lattice hierarchy \cite{metr:AKK}.
Therefore, it is possible to find explicitly the classical solution $\ff$,
playing the role of the background field in the Das--Jevicki
collective theory. Moreover, as we will show,
the integrability manifests itself at the next steps of the derivation,
what allows to complete the analysis of this case.

\subsection{Background field from matrix model solution}

First, we show how the classical solution $\ff$ determining the string
background can be extracted from MQM.
As it was noted, it represents the density of eigenvalues.
The latter is most easily found from the exact form of the Fermi sea
formed by fermions of the MQM singlet sector.
To see how this works,
let us introduce the left and right moving chiral
fields \cite{metr:JEVICKI}:
\be\label{metr:chf}
p_\pm(x,t) = \p_x\Pi \pm \pi \bff(x,t).
\ee
They have the following Poisson brackets
\be\label{metr:Ppp}
\{p_\pm(x),p_\pm(y)\}=\pm 2\pi \p_x\delta(x-y)
\ee
and bring the Hamiltonian to the form
\be\label{metr:Hampp}
H_{\rm coll}=\int {dx\over 2\pi}\, \left( {1\over 6} (p_+^3-p_-^3)+
(u(x)+\mu)(p_+-p_-) \right).
\ee
Using \Ref{metr:Ppp}, it is easy to show that
they should satisfy the Hopf equation
\be\label{metr:Hopfeq}
\p_t p_\pm+p_\pm \p_x p_\pm+\p_x u(x)=0.
\ee
This equation is of integrable type and much easier to solve than
eq. \Ref{metr:Leq}. The key observation, which establishes the contact
with MQM, is that in the quasiclassical limit $p_\pm(x,t)$
determine the upper and lower boundaries of the Fermi sea in
the phase space.
Then the density is given by their difference what
agrees with the definition \Ref{metr:chf}.

For general integrable tachyon perturbations, the solution for
the profile of the Fermi sea has been found in \cite{metr:AKK}.
If we introduce the sources for tachyons with momenta $p_k=k/R$,
$k=1,\dots,n$, it can be represented in the following parametric form
\be\label{metr:SLbac}
\begin{array}{rcl}
p(\o,t)&=&\sum\limits_{k=0}^n a_k
\sinh\left[\left(1-\kR\right)\o+\kR t+\alpha_k\right],\\
x(\o,t)&=&\sum\limits_{k=0}^n a_k
\cosh\left[\left(1-\kR\right)\o+\kR t+\alpha_k \right].
\end{array}
\ee
Here $a_k$ are coefficients which can be found in terms of
the coupling constants $t_k$ of the tachyon operators and the chemical
potential $\mu$ which fixes the Fermi level.
Actually, they depend only on $\mu$ and the products
$\lambda_k^2=t_k t_{-k}$. Instead, the ``phases'' $\alpha_k$
are determined by ratios of the coupling constants.
(Two of them can be excluded by constant shifts of $\o$ and $t$.)
It is easy to check that $p(x,t)$ indeed satisfies
\Ref{metr:Hopfeq} with $u(x)=-\hf x^2$.

\lfig{The Fermi sea of the perturbed MQM.
Its boundary is defined by the two-valued function with two branches
parameterized by $p(\o,t)$ and $\pp(\o,t)$.
The background field $\ff$ coincides with the width of
the Fermi sea.}{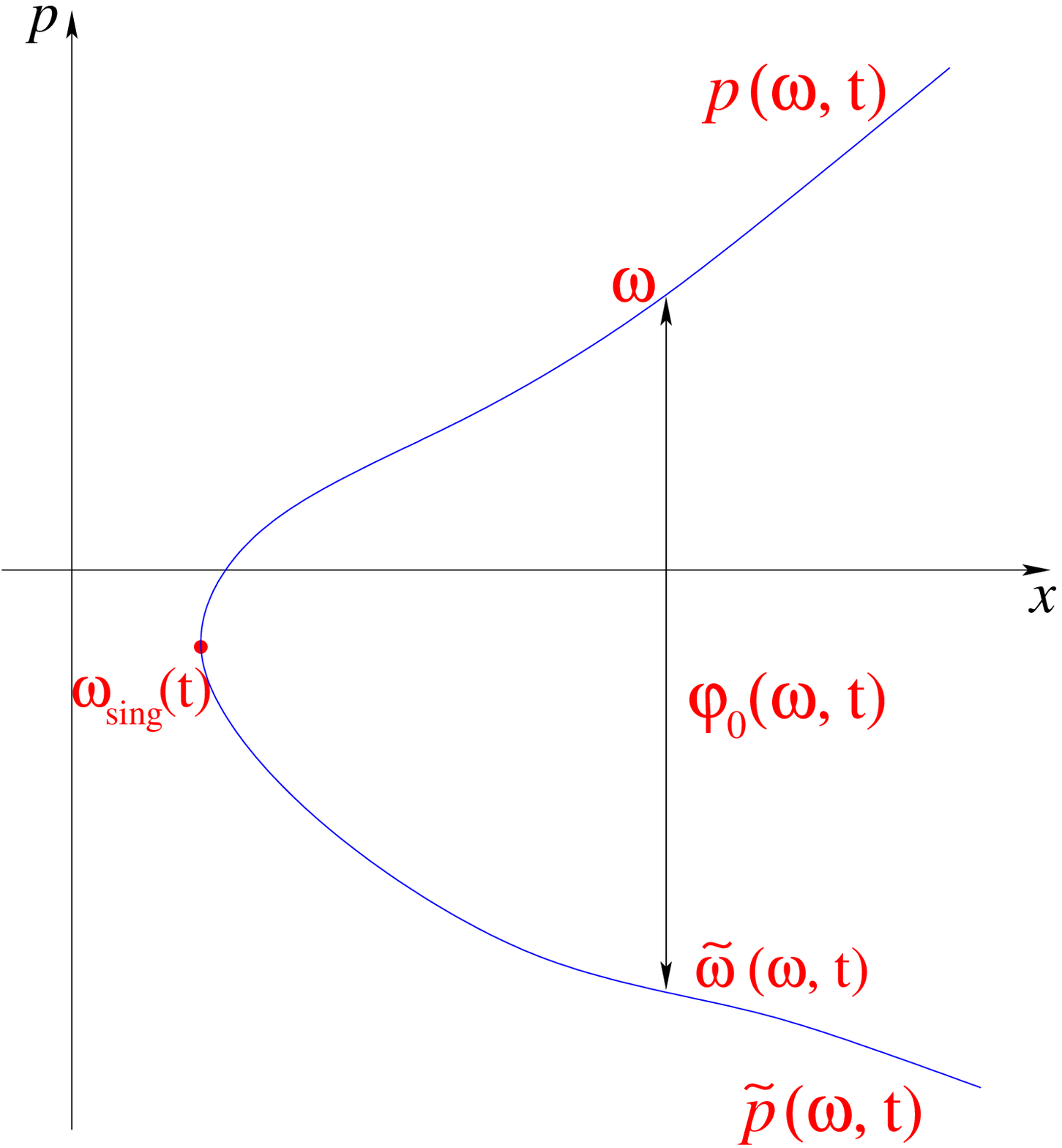}{160pt}{ferms}

To get from this solution $\ff$, note
that $p(x)$ is a two-valued function so that
$p_\pm (x)$ can be identified with its two branches.
As functions of $\o$ they appear as follows.
Let us define a "mirror" parameter $\op(\o,t)$ such that (see fig. 1)
\be\label{metr:ooooo}
x(\op(\o,t),t)=x(\o,t), \ \ \op\ne \o.
\ee
Then $p_+$ can be identified with $p$ and
$p_-$ with $p(\op(\o,t),t)$.
The solution for the background field is given
again in the parametric form
\be\label{metr:solf}
\ff(\o,t)=\hf(p(\o,t)-\pp(\o,t)),
\ee
where we denoted $\pp(\o,t)=p(\op(\o,t),t)$
and $\o$ is related to $x$ by eq. \Ref{metr:SLbac}.
Due to \Ref{metr:cleq}, \Ref{metr:chf} and \Ref{metr:solf},
the effective action \Ref{metr:Sd} can now be rewritten as
\be\label{metr:Sdfpp}
S_{(2)}=\int dt \int {dx\over p-\pp}\, \left[
(\p_t \eta)^2 +  (p+\pp)\p_t\eta \p_x \eta
+p\pp (\p_x\eta)^2  \right].
\ee

\subsection{Flat coordinates}

We know from section 4 that any background
described by the Das-Jevicki effective action
leads to the flat target space metric. However, as we discussed above,
the global structure of the target space can differ
from that of the usual Minkowski space.
To find whether this is the case, one should investigate
the coordinate transformation mapping the target space metric
into the standard form $\eta_{\mu\nu}$.

In general, it is quite a difficult problem to find such a map explicitly.
Remarkably, in our case given by the solution \Ref{metr:SLbac},
this task can be accomplished.
Indeed, we show in Appendix B that the
change of coordinates from $(t,x)$ to $(\tf,\xf)$ with
\be\label{metr:flcor}
\tf=t-{\o+\op\over 2}, \qquad \xf={\o-\op\over 2}
\ee
brings the action \Ref{metr:Sdfpp} to the standard form
\be\label{metr:Sdffl}
S_{(2)}=\pm \hf\int d\tf \int d\xf\,
\left[(\p_\tf\eta)^2-(\p_\xf\eta)^2\right],
\ee
where the sign is determined by the sign of the Jacobian of
the coordinate transformation \Ref{metr:flcor}.
The Jacobian is defined by the function $D$ in (B.2) and can
be represented in the form
\be\label{metr:Jac}
2D= {2(p-\xt)(\pp-\xtp)\over \pp-p },
\ee
where $\xt\equiv\p_t x(\o,t)$ and
$\xtp\equiv\xt(\op,t)$ given explicitly in (B.4).

The zeros of the Jacobian show where the map \Ref{metr:flcor} is degenerate.
All multipliers in \Ref{metr:Jac} vanish at
the line defined by the condition
$\op(\o,t)=\o$ or $\p x/\p \o=0$,
which corresponds to the most left point of the Fermi sea where
two branches of $p(x)$ meet each other (see fig. 1).
In terms of the flat coordinates
\Ref{metr:flcor}, this line is given in the following parametric form
\be\label{metr:singl}
\tf=t-\os(t), \qquad \xf=0,
\ee
where $\os(t)$ is a solution of the above condition.
Also it defines the limiting value of the coordinate $x$, which
should always be larger than $\xs(t)=x(\os(t),t)$. However,
the spacetime can be
analytically continued through this line. This analytical continuation
corresponds to that
in all integrals $x$ should run from $\infty$ to $\xs$ and return back
with simultaneous interchanging the roles of $p$ and $\pp$.
This shows that it is more natural to consider the plane of $(t,\o)$
as the starting point rather than $(t,x)$.
The former is a two-sheet cover of the physical region of the latter.
$\o$ appears as a parameter
along fermionic trajectories and it should not be restricted
to a half-line. Thus, we incorporate the whole scattering picture
and glue two copies of the resulting flat spacetime together
along the line \Ref{metr:singl}.

Unfortunately, there is an ambiguity in the choice of coordinates
where the action has the form \Ref{metr:Sdffl}.  Any conformal change
of variables \Ref{metr:flcor}
\be\label{metr:confch}
(\tf,\xf)\longrightarrow \left(u(t-\o),v(t-\op)\right)
\ee
leaves the action unchanged.  Due to this we can find the form of the
resulting spacetime only up to the conformal map. However, one can
reduce the ambiguity imposing some conditions on the map
\Ref{metr:confch}. First of all, it should be well defined on the
subspace which is the image of the $(t,\o)$-plane under the map
\Ref{metr:flcor}, that is the functions $u$ and $v$ should not have
singularities there.  The next condition is the triviality of
\Ref{metr:confch} in the absence of perturbations since in this limit
\Ref{metr:flcor} reduces to the well known transformation giving the
flat Minkowski space \cite{metr:JEVICKI}.  Besides, we would like to
retain the symmetry $(t,\o)\leftrightarrow (-t,-\o)$ which is explicit
when there are no phases $\alpha_k$ in
\Ref{metr:SLbac}. It reflects the fact the
our spacetime is glued from two identical copies (see discussion in the
previous paragraph). This leads to the condition that $u$ and $v$ are
odd functions.

As a byproduct of our analysis we obtain also
the conformal factor $e^{2\rho}$ appearing in \Ref{metr:mets}.
It should be found, in principle, from
the condition of vanishing curvature for the metric $G_{\mu\nu}$.
Instead, we get it as the Jacobian of the transformation to the
coordinates where $G_{\mu\nu}=\eta_{\mu\nu}$.
If $u$ and $v$ from \Ref{metr:confch} are such coordinates, than
$dt dx \sqrt{-G}=du dv$ and, since $\sqrt{-G}=e^{2\rho}$,
we conclude from \Ref{metr:Jac} that
\be\label{metr:confac}
e^{2\rho}= {\pp-p \over 2(p-\xt)(\pp-\xtp)}u'v'.
\ee
The conformal factor becomes singular on the line \Ref{metr:singl}
discussed above. However, this is only a coordinate singularity
since it disappears after our coordinate transformation.

\subsection{Structure of the target space}

In this subsection we investigate the concrete form of the target
space which is obtained as the image under the coordinate
transformation trivializing the metric. From the previous subsection
it follows that this transformation is a superposition of the map
\Ref{metr:flcor} and a conformal change of coordinates
\Ref{metr:confch}.  Since the latter remains unknown, we will analyze
only the image of the map \Ref{metr:flcor}.  It will be sufficient to
get a general picture because conformal transformations do not change
the causal structure of spacetime.

We restrict our analysis to the case of the simplest non-trivial
perturbation with $n=1$ in \Ref{metr:SLbac}.  It corresponds to the
conformal Sine-Liouville theory. At $R=2/3$ this theory is T-dual to
the CFT suggested to describe string theory on the black hole
background \cite{metr:FZZ}.  We also assume that the coefficients
$a_k$ in \Ref{metr:SLbac} are positive what corresponds to the
positivity of the coupling constants $t_{\pm 1}$.
It is the case which was considered in
\cite{metr:KKK,metr:AK,metr:AKK,metr:AKTBH,metr:AKKNMM}.

To find the image, we investigate the change of coordinates at infinity
of the $(t,\o)$ plane which plays the role of the boundary of the initial
spacetime. In other words, we want to find where the boundary is mapped to.
Thus we assume $t$ and $\o$ to be large and look at the dependence
of $\op$ of their ratio. At the asymptotics $\op(\o,t)$ can be easily
found, but the answer depends non-continuously on $t/\o$ as well as on the
parameter $R$.
There are several different cases which we summarize in Appendix C.

\vskip 0.2cm
$R\ge 1$:
From the results presented in Appendix C it follows
that in this case the image of the map
\Ref{metr:flcor} is the entire Minkowski
spacetime. Thus, we do not see any effect of the tachyon perturbation
on the structure of the target space.

\lfig{Target space of the perturbed theory for
the case $R<1$.}{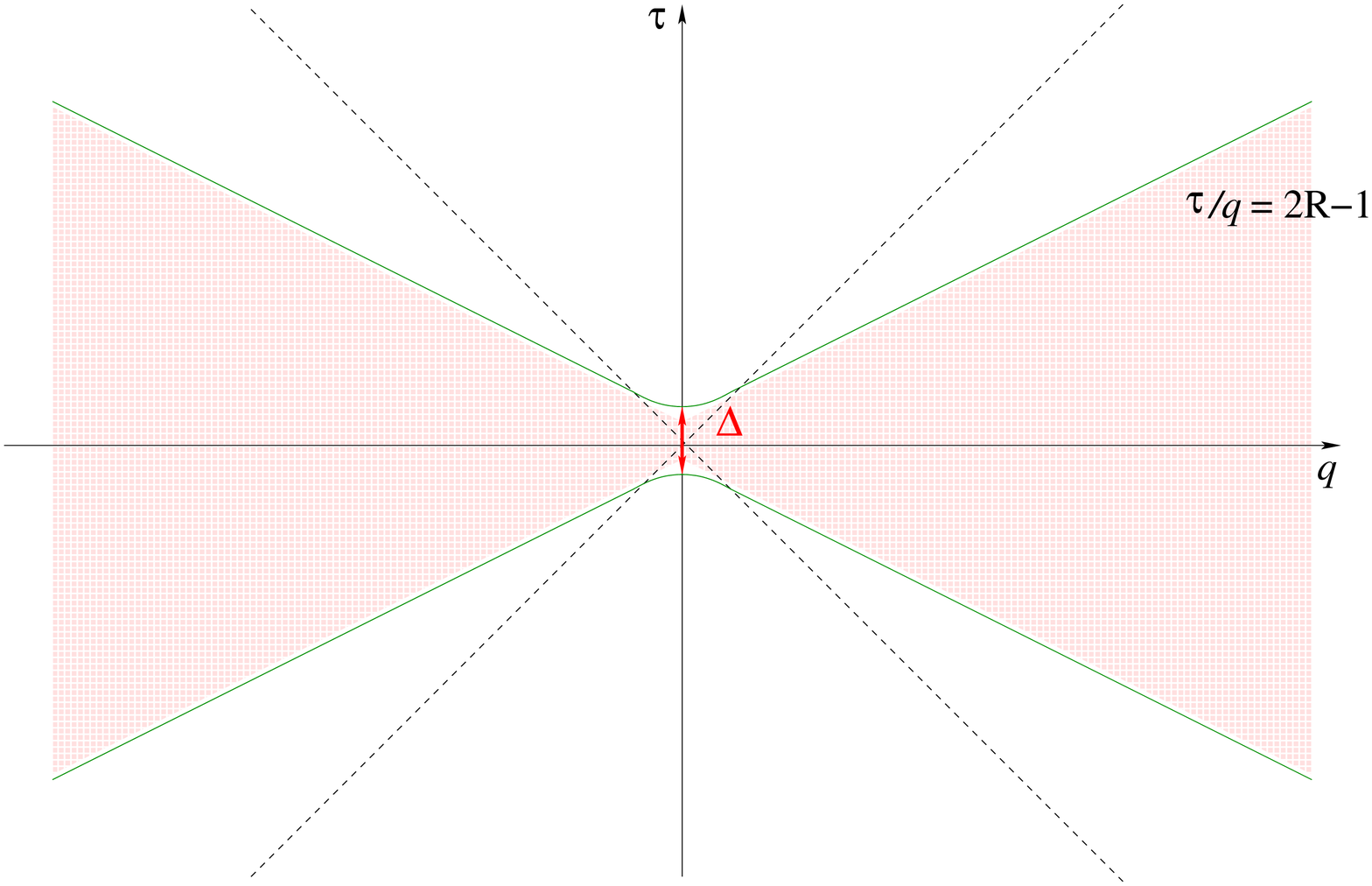}{300pt}{space}

\vskip 0.2cm
$1/2<R<1$:
In this case the resulting spacetime looks like two
conic regions bounded by lines
$\tf=\pm (2R-1)\xf\pm R\log{a_0\over a_1}$ (fig. 2).
Near the origin the boundaries are not anymore the straight lines
but defined by the solution of the following irrational equation
\be\label{metr:oreq}
a_0 e^{\xip}+a_1 e^{(1-1/R)\xip}=a_0 e^{\xi}+a_1 e^{(1-1/R)\xi},
\quad \xip\ne\xi.
\ee
It allows to represent the boundary in the parametric form
\be\label{metr:orbound}
\tf=-\hf(\xi+\xip(\xi)), \qquad \xf=\hf(\xi-\xip(\xi)).
\ee
Two conic regions are glued along some finite interval belonging to
the $\tf$-axis.
It is just that line where the change of coordinates \Ref{metr:flcor}
is degenerate. We can easily find its length.
As it was discussed in the previous subsection, the interval is defined
by eq. \Ref{metr:singl}. We need to find the limiting value of $\tf(t)$ when
$t\rightarrow\infty$.
In this limit the condition for $\os(t)$ can be written as
\be\label{metr:condoo}
2{\p x\over \p \o}\approx a_0 e^{\os}+\left(1-\oR\right)a_1
e^{(1-\oR)\os+\oR t}=0 \Rightarrow \os(t)=t-\hf\Delta,
\ee
where
\be\label{metr:delt}
\Delta={2R} \log\left[{R\over 1-R}{a_0\over a_1}\right].
\ee
Thus $\tf(t)\mathop{\longrightarrow}\limits_{t\rightarrow\infty} \Delta/2$
and we conclude that $\Delta$ is the length we were looking for.

From this result it is easy to understand what happens when we switch off
the perturbation. This corresponds to the limit $a_1\to 0$.
Then the interval $\Delta$ (the minimal distance
between two boundaries) logarithmically diverges and the boundaries
go away to infinity. In this way we recover the entire Minkowski space.
The similar picture emerges in the limit $R\to 1$ when we return to
the previous case $R\ge 1$.

It is interesting to look at the opposite
limit where the length $\Delta$ vanishes.
It happens when $a_1= {Ra_0\over 1-R}$.
The parameters are related to the coupling constant
$\lambda=\sqrt{\tp\tm}$ as follows \cite{metr:AKK}
\be\label{metr:param}
a_0=\sqrt{2}e^{-{1\over 2R}\chi},\qquad
a_1=\sqrt{2}{\lambda}e^{{R-1\over 2R^2}\chi},
\ee
where $\chi=\p^2 \CF/\p \mu^2$ is the second derivative of
the grand canonical free energy.
This quantity is defined by the equation \cite{metr:KKK}
\be\label{metr:freeen}
\mu e^{ {1\over R} \chi} -
\left(1-{1\over R}\right)\lambda^2 e^{{2R-1\over R^2} \chi} =1.
\ee
Taking into account \Ref{metr:param} and \Ref{metr:freeen},
the condition of vanishing $\Delta$ leads to the following
critical point
\be\label{metr:crmu}
\mu_c=-\left(2-{1\over R}\right)\left({1\over R}-1\right)^{1\over 2R-1}
\lambda^{2R\over 2R-1}.
\ee
In \cite{metr:AKK} it was shown that at this point the Fermi sea forms a pinch
and beyond it the solution for the free energy
does not exist anymore. Also this point
coincides with the critical point of Hsu and Kutasov \cite{metr:HSU}, who
interpreted it as a critical point of the pure 2D gravity type.

\vskip 0.2cm
$R=1/2$:
This case can be analyzed using the result (C.1).
We conclude that the spacetime takes the form of a strip bounded
by lines $\tf=\pm \hf\log {a_0\over a_1}$ what agrees also with \Ref{metr:delt}.

\vskip 0.2cm
$R<1/2$:
This interval of values of the parameter $R$ was splited in Appendix C
into two cases. However, they both lead to the same picture.
Moreover, it coincides with one analyzed for $1/2<R<1$ and shown
in fig. 2. Nevertheless, there is an essential distinction with
respect to that case. Using explicit expressions for the new
coordinates from Appendix C, it is easy to check that the Jacobian
of the transformation \Ref{metr:flcor} is negative for $R<1/2$. Therefore,
one has to choose the minus sign in front of the action in \Ref{metr:Sdffl}.
Due to this, the time and space coordinates are exchanged,
so that $\tf$ and $\xf$ are associated with space and time directions,
correspondingly.
As a result, the picture at fig. 2 should now be rotated by $90^{\circ}$.

\subsection{Boundary conditions and global structure}

We found that the introduction of tachyon perturbations in MQM is equivalent
to consider string theory in the flat spacetime of the form
described in the previous subsection (or its conformal transform).
The main feature of this spacetime is the presence of boundaries.
Therefore, to define the theory completely, we should impose
some boundary conditions on the fields propagating there.
We do not see any rigorous way to derive them.
Nevertheless, we discuss the most natural choices for the boundary
conditions and their physical interpretation.

\subsub{Vanishing boundary conditions}

The most natural choice is to choose the vanishing boundary conditions.
It can be justified by that the boundary on fig. 2 came from infinity on
the $(t,\o)$ plane of the matrix model. In the usual analysis of
a quantum field theory one demands that all fields disappear at infinity.
Therefore, it seems to be natural to impose the same condition in our
case also.

The presence of boundaries with vanishing conditions on them can be
interpreted as if one placed the system between two moving mirrors.
However, we note that such interpretation is reliable only for $R<1/2$
where the boundaries are timelike. In the most interesting case
$R\in (1/2,1)$ we would have to suppose that mirrors move faster than
light.

Nevertheless, this interpretation opens an interesting possibility.
It is known that moving mirrors cause particle creation
\cite{metr:BirDav}. In particular, the resulting spectrum can
be thermal so that the system turns out to be at a finite temperature.
On the other hand, in \cite{metr:AKTBH} it was shown that the Sine-Liouville
tachyon perturbation gives rise to the thermal description at $T=2\pi/R$.
Therefore, it would be extremely interesting to reproduce this result
from the exact form of the spacetime obtained here.
In our case the particle creation is a quite expected effect
because two cones lead to the existence of two natural vacua
associated with the right and left cone, correspondingly.
When the modes, defining one of the vacua, propagate from one cone to
another, they disperse on the hole and, in general, will lead to
appearance of particles with respect to the second vacuum.
The concrete form of the particle spectrum depends
crucially on the exact behaviour near the origin since
the creation happens only when the mirror is accelerated
\cite{metr:GMoore,metr:FulDav}.
Unfortunately, in this region
the form of the boundary is known only inexplicitly through
the equation \Ref{metr:oreq}. Besides, the situation is complicated by
the possibility that one needs to make a conformal transform \Ref{metr:confch}
to arrive to the flat coordinates. Usually, such transform can change
thermal properties \cite{metr:DilRev}.
Therefore, it remains unsolved problem
to check whether the particle creation in the found spacetime
reproduces the result obtained in the MQM framework.

\subsub{Periodic conditions}

Although the vanishing boundary conditions are very natural,
it is questionable to apply them in the case of spacelike boundaries.
For example, this would lead to a discrete spectrum
at finite distance from the origin although the spacelike slices
are non-compact. In general, the situation when one encounters
a boundary in time is very strange.

Therefore, in this case it could be more natural to choose
periodic boundary conditions where one identifies the upper boundary
on fig. 2 with the lower one. As a result, the spacetime is compactified.
For $R\in (1/2,1)$ the compact dimension is timelike so that we
find closed timelike curves. Of course, their existence is a very bad
feature. But it might be better than to have spacelike boundaries.
For $R<1/2$ the periodic boundary conditions can also be applied.
Then the spacetime has the topology of a cylinder with a spacelike
compact dimension and we do not encounter the above mentioned problems.
In this case it is also expected that the change of radius
of the cylinder in time will cause particle creation and can
give rise to a temperature.

Note that the picture obtained in this case resembles the orbifold
construction of \cite{metr:NEKR}, where the Minkowski spacetime was factored out
by action of a discrete subgroup of boosts. It leads to four cones,
joint at the origin, corresponding to the causal cones in the
Minkowski spacetime. If one tries to interpret fig. 2 with the
periodic boundary conditions following this analogy, one concludes
that it corresponds to two cones of \cite{metr:NEKR} (spacelike or timelike,
depending on the value of $R$) with the resolved singularity because
the cones are joint not at a point but along some finite interval.

\subsub{Twisted periodic conditions}

There is also another possible choice for the boundary conditions
which comes from comparison with Matrix Quantum Mechanics.
It is very unlikely that this possibility
can be realized but we mention it for completeness.
In MQM to define the density of states and, consequently,
the free energy, one should place the system in a box and also impose
some boundary conditions on the wave functions of fermions
of the singlet sector of MQM.
They relate the scattered (out-going) wave function to the in-coming
one \cite{metr:AKK}. Thus one can say that they identify two infinities
$t,\o\to\infty$ and $t,\o\to -\infty$.
Being applied to our picture, these conditions mean that
one should identify the left lower boundary with the right upper
one and {\it vice-versa}.
In this way we arrive at the compactified spacetime but
where it was ``twisted'' before the compactification.
The topology of the resulting spacetime is that of the M\"obius
sheet.
Note that these boundary conditions have an advantage that
they are invariant under the conformal map \Ref{metr:confch}
if one required from it to preserve the symmetry
$(t,\o)\leftrightarrow (-t,-\o)$.
However, the fact that they lead to a unorienetd spacetime
rises doubts on their credibility.

\vskip 0.2cm

We stop our discussion at this point. We are not able to choose
the right boundary conditions. Due to this, we are not able also to
establish the effect produced by tachyon perturbations on
the target space: either it is the appearance of boundaries or
a compactification. But the form of the obtained spacetime
indicates that one of these modifications of the global structure
does take place.

\ssection{ Conclusions }

We analyzed the proposal that the tachyon perturbations of
2D string theory in the linear dilation background have a dual
description in a non-trivial target space.
Comparing the Das--Jevicki collective field theory with the
low-energy effective action of string theory, we found that
the tachyon perturbations can not change the local structure of
spacetime so that we always remain in the linear dilaton background.
However, the relation between the flat coordinates, where
the background metric has the standard Minkowski form, and the
coordinates of the collective theory coming from Matrix Quantum
Mechanics is affected by the perturbation. As a result,
MQM describes only a part of the Minkowski spacetime.
This introduces boundaries for the target space. But its real
structure depends on the boundary conditions. If they are chosen to be
the periodic conditions, we get a compactified target space.
Unfortunately, we were not able to choose the correct boundary
conditions and, therefore, we can not say what the final form
of the target space is.

One could ask about the physical meaning of the boundaries
because the spacetime can be continued through them to get
the entire Minkowski spacetime.
However, when we work with the tachyon perturbations in the framework
of Matrix Quantum Mechanics, we can not access the parts
of spacetime obtained by this continuation.
The situation can be compared with that of the Unruh effect \cite{metr:Unruh},
when an accelerated observer ``sees'' only the Rindler cone of
the Minkowski spacetime. Thus, we can think of the tachyon perturbations
as if they place the system into another reference frame.
However, there is a crucial difference between the Unruh effect
and our situation. In the former case, the boundaries are lightlike
and considered as event horizons, whereas our boundaries are
spacelike or timelike and one should find for them a physical
interpretation.
In the context of string theory it is tempting to think about them
as branes. Note that spacelike branes (S-branes) have also been
introduced in string theory \cite{metr:Sbranea},
so that the existence of spacelike boundaries should not be an obstacle
for this interpretation. Recently, it was also suggested
a relation between S-branes and thermodynamics \cite{metr:Sbraneb}.

May be one of the most interesting problems, which is left unsolved
in the paper, is to show that the field-theoretic analysis on the
found target space leads to a thermal description with the temperature
$2\pi/R$. This would confirm the result obtained in the MQM framework
\cite{metr:AKTBH} and would be a non-trivial consistency check.
Also, if the periodic boundary conditions are realized, it would be
interesting to elaborate a connection with the work \cite{metr:NEKR}.
This might help to understand how to resolve the orbifold singularity
of the type considered there. In \cite{metr:NEKR} the resolution was found
to be impossible.

Finally, the main problem, which we did not consider here, is
to find the string background corresponding to winding perturbations.
The results obtained in this paper could be useful to approach
the answer if one can realize the world sheet T-duality
directly in the target space. Unfortunately, when the duality
transformation does not act in the direction of a Killing vector,
it is still an open problem.

\bigskip
\noindent{\bf Acknowledgements:}
The author is grateful  to V.~Kazakov, I.~Kostov and D.~Vassilevich
for very valuable discussions.
This work was partially
supported by European Union under the RTN contracts HPRN-CT-2000-00122
and -00131 and by European network EUROGRID HPRN-CT-1999-00161.


\section*{Appendix A. Perturbative solution of dilaton gravity}
\renewcommand{\theequation}{A.\arabic{equation}}
\renewcommand{\thesection}{A.}
\setcounter{equation}{0}

We are interested in the solution of the system of equations \Ref{metr:eqss}
and \Ref{metr:vmas}.
After the substitution $T=e^{\dl}\eta$, they can be rewritten as
\beq
-\hf e^{-\dl}\beta^T-\mt^2\eta &=&
\dd^2\eta=0, \nonumber \\
\mt^2 &=&(\dd\dl)^2-\dd^2\dl-4\pal=0, \label{metr:sfeq} \\
4\alp^{-1}\beta^{\dl}+\beta^G_{\mu\nu}G^{\mu\nu}-4\mt^2
&=&2\dd^2\dl -4\pal e^{2\dl}\eta^2=0,\nonumber \\
\beta^G_{\mu\nu}&=&R_{\mu\nu}+2\dd_{\mu}\dd_{\nu} \dl-
\dd_{\mu}(e^{\dl}\eta) \dd_{\nu}(e^{\dl}\eta)=0. \nonumber
\eeq
Let us choose the conformal gauge
\be\label{metr:cong}
G_{\mu\nu}=e^{2\rho}\eta_{\mu\nu}.
\ee
Then in the light-cone coordinates $\Xpm={X^0\pm X^1\over\sqrt{2}}$
the equations take the following form
\beq
&&\p_+\p_-\eta=0,  \nonumber \\
&&\p_+\p_-\dl- \p_+\dl\p_-\dl=2\pal e^{2\rho} , \nonumber \\
&&\p_+\p_-\dl=-\pal e^{2(\dl+\rho)}\eta^2,
\label{metr:gaugeq}\\
&&\p_+\p_-\rho -\p_+\p_-\dl=-\hf\p_+(e^{\dl}\eta)\p_-(e^{\dl}\eta),\nonumber \\
&&\p_+\p_+\dl=\hf\left(\p_+(e^{\dl}\eta)\right)^2,\nonumber \\
&&\p_-\p_-\dl=\hf\left(\p_-(e^{\dl}\eta)\right)^2,\nonumber
\eeq
where the last three equations are produced by the last equation
in \Ref{metr:sfeq}.

First of all, it is easy to see that if $\eta=0$, \Ref{metr:solld}
is the unique solution.
To generalize this statement, let us consider a solution
which can be represented as an expansion around the linear
dilaton background \Ref{metr:solld}. We use the following ansatz
\be\label{metr:expan}
\rho=\sum_{k>0}\e^{k+1}\rho_k, \qquad
\dl=-\nspal{2}X^1 +\sum_{k>0}\e^{k+1}\ddl_k,
\qquad
\eta=\sum_{k>0}\e^{k}\eta_k.
\ee
In the first order in $\e$ we find
\beq
{(\rm a)}&\ \ &\eta_1=f_1(\Xp)+g_1(\Xm),  \nonumber \\
{(\rm b)}&\ \ &\p_+\p_-\ddl_1 -\sqrt{2\over\alp}(\p_+ - \p_-)\ddl_1=
{4\over \alp}\rho_1 , \nonumber \\
{(\rm c)}&\ \ &\p_+\p_-\ddl_1=
-{1\over\alp} e^{-\nspal{4}X^1}\eta_1^2,
\label{metr:firsto}  \\
{(\rm d)}&\ \ &\p_+\p_-(\rho_1 -\ddl_1)=-\hf
\p_+\left(e^{-{\nspal{2}}X^1}\eta_1\right)
\p_-\left(e^{-{\nspal{2}}X^1}\eta_1\right),\nonumber \\
{(\rm e)}&\ \ &\p_+\p_+\ddl_1=\hf
\left(\p_+\left(e^{-{\nspal{2}}X^1}\eta_1\right)\right)^2,\nonumber \\
{(\rm f)}&\ \ &\p_-\p_-\ddl_1=\hf
\left(\p_-\left(e^{-{\nspal{2}}X^1}\eta_1\right)\right)^2. \nonumber
\eeq
Using equations (b) and (c), one can check that (d) is fulfilled identically.
On the other hand, for (c), (e) and (f) to be consistent, some
integrability conditions must hold. They are obtained comparing
the derivatives of these equations. Taking into account (a),
we get the following
conditions on $f_1(\Xp)$ and $g_1(\Xm)$
\be\label{metr:condfg}
\begin{array}{c}
(f'-g')\left(f'-\sqrt{2\over\alp}(f+g)\right) =0, \\
(f'-g')\left(g'+\sqrt{2\over\alp}(f+g)\right) =0.
\end{array}
\ee
The only solution of \Ref{metr:condfg} is obtained when $f'=g'$.
Then we find
\be\label{metr:solfirst}
\rho_1={\alp c_1^2\over 32}e^{-{\nspal{4}}X^1} , \quad
\ddl_1={1\over 8} \left( (c_1 X^0+d_1)^2+{\alp c_1^2\over 4}\right)
e^{-{\nspal{4}}X^1}, \quad
\eta_1=c_1 X^0+d_1,
\ee
where $c_1$ and $d_1$ are some constants.
In the second order we get the similar system of equations
as \Ref{metr:firsto},
where one should change one of $\eta_1$ in terms quadratic
in tachyon by $2\eta_2$.
Then we obtain
\beq
&\rho_2={\alp c_1 c_2\over 16}e^{-{\nspal{4}}X^1} , \quad
 \ddl_2={1\over 4} \left( (c_1 X^0+d_1)(c_2 X^0+d_2)+
{\alp c_1 c_2\over 4}\right)
e^{-{\nspal{4}}X^1}, & \nonumber \\
& \eta_2=c_2 X^0+d_2. & \label{metr:solsec}
\eeq
But already in the next order in $\e$ we encounter inconsistences
which can not be overcomed.
We must choose the vanishing tachyon and all other perturbations.
Thus, we conclude that the only solution of the system \Ref{metr:sfeq},
which can be represented in the form \Ref{metr:expan},
is the linear dilaton background.

\section*{Appendix B. Change to the flat coordinates}
\renewcommand{\theequation}{B.\arabic{equation}}
\renewcommand{\thesection}{B.}
\setcounter{equation}{0}

In this appendix we demonstrate
that the quadratic part of the effective action for the tachyon field
\Ref{metr:Sdfpp} in the coordinates \Ref{metr:flcor} takes the form \Ref{metr:Sdffl}.

First of all, for generic functions $x$, $p$ and $\op$ of $\o$ and $t$
the passage to the variables \Ref{metr:flcor} leads to the action
\be\label{metr:Sdint}
S_{(2)}=\hf\int d\tf \int d\xf\, {|D|\over p-\pp}
\left[A(\p_\tf\eta)^2-2B\p_\tf\eta \p_\xf\eta-C(\p_\xf\eta)^2\right],
\ee
where
\beq
A&=&\left[ 2-{\p_t\op}-\left({\p_\o x}\right)^{-1}
\left( 1+{\p_\o \op} \right)\left( p-{\p_t x}\right)\right]
\left[ 2-{\p_t\op}-\left({\p_\o x}\right)^{-1}
\left( 1+{\p_\o \op} \right)\left(\pp-{\p_t x}\right)\right],
\nonumber \\
B&=&{\p_t \op}\left( 2-{\p_t \op}\right)-
\left({\p_\o x}\right)^{-1}\left(p+\pp-2{\p_t x}\right)
\left( 1-{\p_\o \op}+{\p_t \op}{\p_\o \op} \right)
\nonumber \\
&+&\left({\p_\o x}\right)^{-2}
\left( 1-{\p_\o \op} \right)\left( 1+{\p_\o \op} \right)
\left(p-{\p_t x}\right)\left(\pp-{\p_t x}\right),
\label{metr:coefff} \\
C&=&\left[{\p_t\op}-\left({\p_\o x}\right)^{-1}
\left( 1-{\p_\o \op} \right)\left( p-{\p_t x}\right)\right]
\left[ {\p_t\op}-\left({\p_\o x}\right)^{-1}
\left( 1-{\p_\o \op} \right)\left(\pp-{\p_t x}\right)\right], \nonumber \\
D&=& {{\p_\o x} \over 1-\left({\p_\o\op}+{\p_t\op}\right)}.
\nonumber
\eeq
From the explicit form \Ref{metr:SLbac} of $p(\o,t)$ and $x(\o,t)$ it is easy to
check the following properties
\be\label{metr:porper}
{\p x\over \p t}=p-{\p x\over \p \o}, \qquad
{\p \op\over \p \o}={{\p x/\p\o} \over \pp -\xtp}, \qquad
{\p \op\over \p t}={p-\xtp \over \pp -\xtp}-{\p \op\over \p \o}
\ee
with
\be\label{metr:dobav}
\xtp\equiv {\p_t x}(\op,t)=\oR\sum\limits_{k=1}^n k a_k
\sinh\left[(1-\kR)\op+\kR t\right].
\ee
These properties are enough to show that
$DA=DC=p-\pp$ and $B=0$ what gives the action \Ref{metr:Sdffl}.

\section*{Appendix C. Asymptotic values of the coordinates}
\renewcommand{\theequation}{C.\arabic{equation}}
\renewcommand{\thesection}{C.}
\setcounter{equation}{0}

\def\vdist{\vskip 0.2cm}
\def\vdst{\nobreak  \vskip 0.05cm  \nobreak}

\def\tbcol#1{\vrule\quad\hfil#1\hfil\quad}
\def\tblh{&\omit\vrule height2pt&&&\cr}
\def\tbline{
\tblh
\noalign{\hrule}
\tblh
}

In this appendix we display the results for $\op$, $\tf$ and $\xf$
as functions of $t$ and $\o$ which are obtained in the asymptotics
$t,\o\rightarrow \infty$ in the case of Sine--Liouville theory ($n=1$).
The results depend on the ratio $t/\o$ as well as on the radius parameter.
There are four different cases dependent of the value of $R$.
We summarize them in four tables.
In each table we distinguish several regions of $t/\o$.
For each region we give the asymptotic expressions for the coordinates,
the corresponding interval of values of $\tf/\xf$, and the coefficients
in front of the constant term $\log{a_0\over a_1}$,
which should be added to $\op$.
The latter are given for $\o>0$. For $\o<0$ one should change their sign.

\vdist

$
R>1
$

\vdst

\vbox{\offinterlineskip
\hrule
\halign{&\tbcol{#}&\tbcol{#}&\tbcol{#}&\tbcol{#}&\vrule# \cr
\tblh
$t/\o$ & $(-1,1)$ & $(1,\infty)\cup(-\infty,1-2R)$ & $(1-2R,-1)$  & \cr
\tbline
$\op$ & $-\o$ &
$-\left(1-\oR\right)\o-\oR t $ &  
$-{1\over R-1}\left(R\o+t\right)$ & 
\cr
$\tf$ & $t$ &
$\left(1+{1 \over 2R}\right)t-{1\over 2R}\o$ & 
${1 \over 2(R-1)}\left((2R-1)t+\o\right)$ & 
\cr
$\xf$ & $\o$ &
$\left(1-{1 \over 2R}\right)\o+{1\over 2R}t$ & 
${1 \over 2(R-1)}\left((2R-1)\o+t\right)$ & 
\cr
$\tf/\xf$ & $(-1,1)$ & $(1,2R+1)\cup(2R+1,\infty) $ & $(-\infty,-1)$ &   \cr
$\log{a_0\over a_1}$ & $0$ & $\pm 1$ & $-{R \over R-1}$ &   \cr
\tblh
\noalign{\hrule}\cr
}}

\vdist

$
1/2<R<1
$

\vdst

\vbox{\offinterlineskip
\hrule
\halign{&\tbcol{#}&\tbcol{#}&\tbcol{#}&\tbcol{#}&\vrule# \cr
\tblh
$t/\o$ & $(1-2R,2R-1)$ & $(2R-1,1)$ & $(1,\infty)\cup(-\infty,1-2R)$  & \cr
\tbline
$\op$ & $-\o$ &
$-{1\over 1-R}\left(R\o- t\right) $ & 
$-\left(\oR-1\right)\o+\oR t$ & 
\cr
$\tf$ & $t$ &
${2R-1\over 2(1-R)}\left(\o-t\right)$ & 
$\left(1-{1\over 2R}\right)\left(t-\o\right)$&
\cr
$\xf$ & $\o$ &
${1 \over 2(1-R)}\left(\o-t\right)$ & 
${1 \over 2R}\left(\o-t\right)$ &
\cr
$\tf/\xf$ & $(1-2R,2R-1)$ & $\approx (2R-1) $ & $\approx(1-2R)$ &   \cr
$\log{a_0\over a_1}$ & $0$ & $-{R\over 1-R}$ & $\mp 1$ &   \cr
\tblh
\noalign{\hrule}\cr
}}

\vdist

$
1/3<R<1/2
$

\vdst

\vbox{\offinterlineskip
\hrule
\halign{&\tbcol{#}&\tbcol{#}&\tbcol{#}&\tbcol{#}&\vrule# \cr
\tblh
$t/\o$ & $({(1-R)(1-2R)\over1-3R},1-2R)$  &
$(1-2R,1)$ & $(1,\infty)\cup(-\infty,{(1-R)(1-2R)\over1-3R})$ & \cr
\tbline
$\op$ &
$-\o+{2\over 1-R}t $ &
$-{1\over 1-R}\left(R\o- t\right)$ & 
$-\left(\oR-1\right)\o+\oR t$ & 
\cr
$\tf$ &
$-{R\over 1-R}t$ &
${2R-1\over 2(1-R)}\left(\o-t\right)$ & 
$\left(1-{1\over 2R}\right)\left(t-\o\right)$&
\cr
$\xf$ & $\o-{1\over 1-R}t$ &
${1 \over 2(1-R)}\left(\o-t\right)$ & 
${1 \over 2R}\left(\o-t\right)$ &
\cr
$\tf/\xf$ & $(2R-1,1-2R)$ & $\approx (2R-1) $ & $\approx(1-2R)$ &   \cr
$\log{a_0\over a_1}$ & $0$ & $-{R\over 1-R}$ & $\mp 1$ &   \cr
\tblh
\noalign{\hrule}\cr
}}

\vdist

$
R<1/3
$

\vdst

\vbox{\offinterlineskip
\hrule
\halign{&\tbcol{#}&\tbcol{#}&\tbcol{#}&\tbcol{#}&\vrule# \cr
\tblh
$t/\o$ & $({(1-R)(1-2R)\over1-3R},\infty)\cup(-\infty,1-2R)$  &
$(1-2R,1)$ & $(1,{(1-R)(1-2R)\over1-3R})$ & \cr
\tbline
$\op$ &
$-\o+{2\over 1-R}t $ &
$-{1\over 1-R}\left(R\o- t\right)$ & 
$-\left(\oR-1\right)\o+\oR t$ & 
\cr
$\tf$ &
$-{R\over 1-R}t$ &
${2R-1\over 2(1-R)}\left(\o-t\right)$ & 
$\left(1-{1\over 2R}\right)\left(t-\o\right)$&
\cr
$\xf$ & $\o-{1\over 1-R}t$ &
${1 \over 2(1-R)}\left(\o-t\right)$ & 
${1 \over 2R}\left(\o-t\right)$ &
\cr
$\tf/\xf$ & $(R,1-2R)\cup(2R-1,R)$ & $\approx (2R-1) $ & $\approx(1-2R)$ &
\cr
$\log{a_0\over a_1}$ & $0$ & $-{R\over 1-R}$ & $- 1$ &   \cr
\tblh
\noalign{\hrule}\cr
}}

\def\tblhsm{&\omit\vrule height2pt&&\cr}
\def\tblinesm{
\tblhsm
\noalign{\hrule}
\tblhsm
}

\vdist

Finally, there are also two critical cases $R=1$ and $R=\hf$.
In these cases one can find explicitly the transformation of coordinates
\Ref{metr:flcor} on the entire plane. The results are
\be
\label{metr:Rcas}
\begin{array}{ccc}
 R=1 &\qquad & R=1/2 \\
\op=-\o &\qquad & \op=-\o+\log{a_0+a_1 e^{2t}\over a_0+a_1 e^{-2t}}
\\
\tf=t &\qquad & \tf=t-\hf\log{a_0+a_1 e^{2t}\over a_0+a_1 e^{-2t}}
\\
\xf=\o &\qquad & \xf=\o-\hf\log{a_0+a_1 e^{2t}\over a_0+a_1 e^{-2t}}
\end{array}
\ee

\end{document}